\documentclass[10pt,conference]{IEEEtran}
\usepackage{cite}
\usepackage{xspace}
\usepackage[pdftex]{graphicx}
\usepackage{url}
\usepackage{subcaption}
\usepackage{booktabs}
\usepackage[linesnumbered,lined,ruled,boxed]{algorithm2e}
\usepackage{multirow}
\usepackage{makecell}
\usepackage{MnSymbol}
\usepackage{diagbox}
\usepackage[skins]{tcolorbox}
\usepackage{hyperref}

\usepackage{wrapfig}
\usepackage{picins}

\newcommand{\smart}[1][]{\textsc{BFT-SMaRt}\xspace}
\newcommand{\smartchain}[1][]{\textsc{SMaRtChain}\xspace}

\begin{document}
\title{From Byzantine Replication to Blockchain:\\
Consensus is only the Beginning}

\author{
Alysson Bessani\IEEEauthorrefmark{1}, Eduardo Alchieri\IEEEauthorrefmark{2}, Jo\~ao Sousa\IEEEauthorrefmark{1}, Andr\'e Oliveira\IEEEauthorrefmark{1}, Fernando Pedone\IEEEauthorrefmark{3} \\
{\small \IEEEauthorrefmark{1}LASIGE, Faculdade de Ci\^encias, Universidade de Lisboa, Portugal} \\
{\small \IEEEauthorrefmark{2}Departamento de Ci\^encia da Computa\c{c}\~ao, Universidade de Bras\'ilia, Brasil}\\
{\small \IEEEauthorrefmark{3}Universit\`{a} della Svizzera Italiana, Lugano, Switzerland}
}

\maketitle


\begin{abstract}
The popularization of blockchains leads to a resurgence of interest in Byzantine Fault-Tolerant (BFT) state machine replication protocols.
However, much of the work on this topic focuses on the underlying consensus protocols, with emphasis on their lack of scalability, leaving other subtle limitations unaddressed.
These limitations are related to the effects of maintaining a durable blockchain instead of a write-ahead log and the requirement for reconfiguring the set of replicas in a decentralized way.
We demonstrate these limitations using a digital coin blockchain application and \smart, a popular BFT replication library.
We show how they can be addressed both at a conceptual level, in a protocol-agnostic way, and by implementing \smartchain, a blockchain platform based on \smart.
\smartchain improves the performance of our digital coin application by a factor of eight when compared with a naive implementation on top of \smart.
Moreover, \smartchain achieves a throughput $8\times$ and $33\times$ better than Tendermint and Hyperledger Fabric, respectively, when ensuring strong durability on its blockchain.\footnote{This is a preprint of a paper to appear on the 50th IEEE/IFIP Int. Conf. on Dependable Systems and Networks (DSN'20).}
\end{abstract}


\section{Introduction}

Recent years have seen a resurgence of interest in state machine replication (SMR) protocols, specifically in the context of permissioned blockchain systems~\cite{Androulaki2018,chain2017,Buchman2016,Martino2016}.
Such protocols are used to maintain a set of stateful replicas, which execute the same set of requests in the same order, deterministically. 
Byzantine Fault-Tolerant (BFT) state machine replication protocols such as PBFT~\cite{Cas99} and its descendants~\cite{Cow06,Kot07,Amir2011,Ver09,Veronese13,Aublin2013} are particularly relevant, as they implement the model properties even in the presence of an adversary that may be able to corrupt and control a fraction of the replicas.
Such protocols are a direct fit for permissioned blockchains~\cite{cachin17}, where every peer/replica is known and approved to participate in the system.
They are also a fundamental building block for some recent high-performance permissionless or open blockchains (e.g.,~\cite{Pass2017,Kokoris-Kogias2016,abraham2017,Yu2019}) that elect a subset of peers to be a transaction processing committee running the BFT protocol.

Most of the recent research on BFT replication applied to blockchain has focused on the scalability of the underlying consensus protocol~\cite{Miller2016,Duan2018,Gilad2017,Golan-Gueta2019,Yang2018,Liu2019,Crain18,Guerraoui2019}, as most BFT protocols described before were typically designed considering few replicas.
Nevertheless, there are other subtle but important differences among the BFT state machine replication approach and blockchains.
While many replicated state machine protocols build an internal log of executed operations for state synchronization after a leader change or a replica recovery, a blockchain system differs from traditional SMR as such log must (1) be written to stable storage to ensure durability, (2) include the result of the transactions for auditing purposes, and (3) be self-verifiable by any third party.
Another key difference is that while the vast majority of the literature about BFT SMR assumes a static set of processes, in a blockchain consortium, peers are expected to join and leave at any time, without the need for an additional trusted party.

In this paper, we show that these differences lead to inherent limitations, which we demonstrate by designing and running a simple digital coin blockchain application on top of \smart~\cite{Bessani2014}, a well-known BFT replication library. Our experiments show that depending on how the blockchain is implemented, and how much we are willing to trade in terms of blockchain features for better integration with the SMR library, the system throughput can go from 1.7k to 14.8k txs/sec.

Furthermore, we identify subtle issues related with \emph{transactions persistence} and \emph{blockchain forks}.
More specifically, we show that it is possible to lose a suffix of the committed transaction history in case of a full crash of the system.
This calls into question the finality of permissioned blockchains and makes them weaker in terms of durability than the centralized transactional systems they are supposed to replace.
Additionally, we observe that blockchain forks might appear as a side effect of run-time consortium reconfigurations since compromised keys from past members of the consortium can be used to generate such forks.

We show that these limitations can be addressed at a conceptual level in a \emph{protocol-agnostic way}, by describing novel mechanisms for \emph{efficiently logging transactions and their results as a self-verifiable chain of immutable blocks} and \emph{reconfiguring the replica set in a secure and decentralized way}.
These mechanisms are independent from the consensus protocol employed to order transactions, being thus general enough to be potentially useful for any blockchain system.

The proposed techniques were implemented on \smartchain, a blockchain platform based on \smart. 
\smartchain improves the performance of the digital coin application by a factor of $8$ when compared with running it on top of \smart, and provides a performance $8\times$ and $33\times$ better than existing comparable production-level blockchains like Tendermint~\cite{Buchman2016} and Hyperledger Fabric~\cite{Androulaki2018}, respectively.

In summary, this paper makes the following contributions:

\begin{enumerate}

\item It identifies three fundamental limitations of running blockchain applications on top of ``classical'' BFT SMR protocols: one related with potential performance issues, and two related with the gap between the state machine replication approach and blockchain requirements;

\item It introduces solutions for addressing these limitations, namely: an efficient design for transforming SMR logs in blockchains, a protocol for increasing the durability guarantee of the system, and new strategies for reconfiguring the replica set without opening breaches for blockchain forks;

\item It describes \smartchain, an experimental permissioned blockchain platform corresponding to the implementation of these techniques, and its evaluation showing it achieves significant performance gains when compared with similar systems.

\end{enumerate}

The remainder of this paper is organized as follows.
Section~\ref{sec:background} presents the relevant background on blockchain and state machine replication, including \smart.
Section~\ref{sec:system-model} presents our system and adversary model. 
The gap between the SMR and blockchain is discussed in Section~\ref{sec:limitations}.
The \smartchain platform is described in Section~\ref{sec:smartchain}.
Section~\ref{sec:evaluation} presents the experimental evaluation of \smartchain.
Finally, some related works and concluding remarks are presented in Sections~\ref{sec:rel-work} and~\ref{sec:conclusions}, respectively.

\section{Background}
\label{sec:background}


\subsection{Blockchain}
\label{sec:blockchain}

The concept of blockchain was introduced by Bitcoin to solve the double spending problem associated with cryptocurrencies in open peer-to-peer networks~\cite{Nakamoto_bitcoin:a}.
A blockchain is an open database that maintains a distributed ledger comprised by a growing list of records called \emph{blocks}, each of them containing transactions executed by the system.
This authenticated data structure~\cite{Tamassia2003}
consists of a sequence of blocks in which each one contains the cryptographic hash of the previous block in the chain.
This ensures that block $j$ cannot be forged without also forging all subsequent blocks $j+1...i$.


A distributed system implements a \emph{robust transaction ledger} (i.e., a blockchain) if it satisfies the following two properties (adapted from~\cite{Garay2015}):

\begin{itemize}

\item \emph{Persistence}: If a correct node reports a ledger that contains a transaction $tx$ in a block more than $k$ blocks away from the end of the ledger, then $tx$ will eventually be reported in the same position in the ledger by any honest node of the system.

\item \emph{Liveness}: If a transaction is provided as input to all correct nodes, then there exists a correct node who will eventually report this transaction at a block more than $k$ blocks away from the end of the ledger.

\end{itemize}

Blockchain systems satisfy these properties abiding to either the \emph{permissionless} or \emph{permissioned} models~\cite{Vukolic2016}.
Permissionless blockchains are maintained across peer-to-peer networks in a completely decentralized and anonymous manner~\cite{Nakamoto_bitcoin:a, Wood15}.
In order to determine the next block to append to the ledger, peers need to execute a Proof-of-Work (PoW) to create a valid block~\cite{Garay2015} (or an equivalent mechanism, e.g., Proof-of-Stake~\cite{Gilad2017,Kiayias2017}) that is then disseminated to the network.
The key idea behind the permissionless consensus, employed in Bitcoin and Ethereum,  is to prevent an adversary from creating new blocks faster than honest participants.
The first participant that finds such a solution gets to append its block to the ledger on all correct peers.
Therefore, intuitively, as long as the adversary controls less than half of the total computing power present in the network, it is unable to tamper with the ledger.\footnote{In fact the speed of the network also affects the maximum adversarial power tolerated, which is typically assumed to be much smaller than 50\%~\cite{Garay2015}.}
This phenomenon also enables participants to establish a total order on the transactions by adopting the longest ledger with a valid PoW as the \emph{de facto} transaction history.

The PoW mechanism makes permissionless blockchains slow and extremely energy demanding~\cite{Vukolic2016}.
By contrast, permissioned blockchains do not expend as many resources and are able to reach better transaction latency and throughput.
This is because nodes participating in this type of ledgers execute a traditional BFT consensus (e.g., PBFT~\cite{Cas99}) to decide on the next block to be appended to the ledger~\cite{cachin17}.
However, this approach requires a consortium of nodes that know each other for executing the consensus protocol.
In this scenario, the bound on the adversary's power is structural, not computational, i.e., safety is ensured as long as the adversary controls less than a fraction of the nodes (usually a third).


\subsection{State Machine Replication}
\label{sec:smr}

In the state machine replication approach~\cite{Lam78,Sch90}, an arbitrary number of client processes issue requests to a set of replicas.
These replicas implement a stateful service that receives these requests and updates its state accordingly to the operation contained in the clients' requests.
Once enough replicas transmit matching replies to the client, its invocation returns the result computed by the service.

The goal of this technique is to make the service state maintained by each replica evolve in a consistent way.
In order to achieve this behavior, it is necessary to satisfy the following requirements~\cite{Sch90}:

\begin{enumerate}

\item Any two correct replicas $r$ and $r'$ start with state $s_0$;

\item If any two correct replicas $r$ and $r'$ apply operation $o$ to state $S$, both $r$ and $r'$ will obtain state $S'$;

\item Any two correct replicas $r$ and $r'$ execute the same sequence of operations $o_0,...,o_i$.

\end{enumerate}

The first two requirements can be easily fulfilled if the service is deterministic, but the last one requires a \emph{total order broadcast} primitive, which is equivalent to solving the \emph{consensus problem}~\cite{Hadzilacos93}.








\subsection{The \smart Library}
\label{sec:bftsmart}

\smart~\cite{Bessani2014} is an open-source library that implements a modular SMR protocol~\cite{Sousa12}, as well as features such as state transfer and group reconfiguration.
In this section we describe these features as they are fundamental for any practical deployment of SMR.

\subsubsection{SMR protocol}
\label{sec:bftsmart-modsmart}









\begin{figure}[t]
	\centering
        \includegraphics[page=1,scale=0.9]{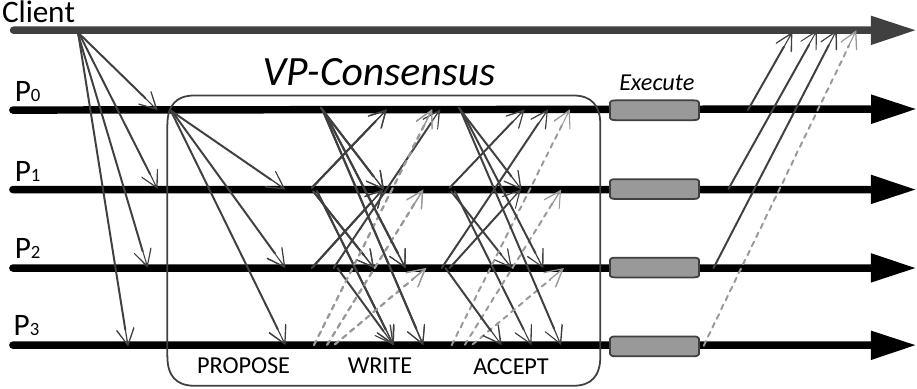}
    \caption{\smart ordering message pattern.}
    \label{fig:normal-pattern}
\end{figure}

\smart uses the Mod-SMaRt protocol to implement the SMR properties described in Section~\ref{sec:smr}.
Mod-SMaRt is a modular SMR protocol that works by executing a sequence of consensus instances based on the BFT consensus algorithm described in~\cite{Cac09}.
During normal operation, the resulting communication pattern is similar to the well-known PBFT protocol~\cite{Cas99} (Figure \ref{fig:normal-pattern}).
Each consensus instance $i$ begins with a leader replica proposing a batch of client operations to be decided within that instance.
All replicas that receive the proposal verify if its sender is correct by exchanging $\texttt{WRITE}$ messages containing a cryptographic hash of the proposed batch with all other replicas.
If a replica receives $\texttt{WRITE}$ messages with the same hash from more than two thirds of the replicas, it sends a signed $\texttt{ACCEPT}$ message to all others containing this hash.
If a replica receives $\texttt{ACCEPT}$ messages for the same hash from more than two thirds of the replicas, it delivers the corresponding batch as the decision for this consensus instance, alongside \emph{a proof comprised by the set of signed messages} received in this last phase.

If the leader replica is faulty and/or the network experiences a period of asynchrony, Mod-SMaRt may trigger a \emph{synchronization phase} to elect a new leader for the consensus instances and synchronize all correct replicas~\cite{Sousa12}.

\subsubsection{State transfer}
\label{sec:bftsmart-statetransfer}

\smart also allows crashed replicas to recover and resume execution.
This is done by using an intermediate layer between the Mod-SMaRt protocol and the replicated service, which is responsible for triggering service checkpoints and managing the request log.

The library provides two state transfer implementations in this layer.
One uses an approach similar to PBFT that consists of storing the request log in memory which is periodically truncated after a snapshot of the service state is created.
A recovering replica obtains the state by probing other replicas about their last completed consensus instance and asking $f+1$ replicas to send the version of the state up to that instance.\footnote{In order to render this mechanism as efficient as possible, only one replica sends the entire state, while other $f$ replicas send only a hash of it~\cite{Cas99}.}

The other implementation is the durability layer described in \cite{Bessani13}.
When this layer is enabled, \smart stores the request log into stable storage to preserve the service state even if all replicas fail by crashing.
In order to write requests to disk as efficiently as possible, delivered requests are written to the durable log in parallel with their execution by the service.
To better exploit the large bandwidth of stable storage devices, the system tries to write multiple batches at once, diluting the cost of a synchronous write among many requests. 
More specifically, the latency of writing one or ten request batches in the stable log is similar, yet the throughput would ultimately increase roughly by a factor of 10 in the latter~\cite{Bessani13}.

This durability layer also enables replicas to execute checkpoints at different moments of their execution and a collaborative state transfer.
These features alleviate the performance degradation caused by checkpoint generation and state transfer when the system is under heavy load.


\subsubsection{Group Reconfiguration}
\label{sec:bftsmart-reconfig}

\smart provides mechanisms for reconfiguring the replica set.
In particular, the reconfiguration mechanism assumes the existence of a distinguished trusted client known as the \emph{View Manager}, which uses the aforementioned state machine protocol to issue updates to the replica set.
To change the current replica set (\emph{view}) of the system, the View Manager issues a signed reconfiguration request that is submitted just like any other client operation.
However, this request is never delivered to the application and instead is used to update the view.
Since these special operations are also totally ordered, all replicas will observe the same updates to the view along the system's lifespan.

Once the View Manager receives confirmation from the current replicas that its update was executed, it notifies the joining replicas that they can start participating in the replication protocol.
At this point, they invoke the state transfer protocol to retrieve the latest application state from other replicas (as described previously) before actively participating in the replication protocol.
Once these replicas receive and install the state, they are ready to process new requests.

\section{System Model}
\label{sec:system-model}

We consider a fully-connected distributed system composed by a universe of processes $U$ that can be divided in two subsets: an infinite set of replicas $\Pi = \{r_1,r_2,...\}$, and an infinite set of clients $C = \{c_1,c_2,...\}$.
Clients access the blockchain/SMR system maintained by a subset of the replicas (a \emph{view}) by sending their transactions to be executed and appended to the blockchain maintained by these replicas.
Each process (client or server) of the system has a unique identifier.
Servers and clients are prone to \emph{Byzantine failures}.
Byzantine processes are said to be \emph{faulty}.
A process that is not faulty is said to be \emph{correct}. 
Each process has a \emph{permanent public-private key pair} and has access to cryptographic functions for digital signatures and secure hashes.
We assume all processes can obtain the public keys of other processes by standard means.
Moreover, there are authenticated fair point-to-point links connecting every pair of processes. 

We assume further an eventually synchronous system  model~\cite{Dwork1988}.
This means the network may behave asynchronously until some unknown instant $T$ after which it becomes synchronous, i.e., time bounds for computation and communication shall be enforced after $T$.


\paragraph{Dynamic replica groups}
During system execution, a sequence of views is installed to account for replicas joining and leaving.
Process arrivals follow the infinite arrival model with bounded (and unknown) concurrency~\cite{Aguilera2004}.
We assume a non-empty initial view $v_\mathit{init}$ known to all processes (e.g., which is written in the genesis block, as will be discussed in later sections).
The system \emph{current view} $cv$ represents the most up-to-date view installed in the system, with its replicas being the only ones that may participate in the execution of the ordering protocol.
We denote by $cv.n$ the number of replicas in $cv$ and $cv.f$ the number of replicas in $cv$ allowed to fail, being $cv.f \leq \lfloor \frac{cv.n-1}{3}\rfloor$.
A replica that asks to leave the system must remain executing the protocols until it knows that a more up-to-date view is installed, otherwise it is considered faulty.

\paragraph{Crashes and recoveries}
We consider that \emph{all replicas} in $cv$ are subject to \emph{recoverable crashes}, i.e., all replicas can crash at once. 
A replica that is in the process of being restarted is said to be in \emph{recovery mode} and cannot participate in the ordering protocol until its service state is restored.
Therefore, the system only make progress when there are at most $cv.f$ faulty and recovering replicas.

In order to potentially bring back the entire set of replicas in $cv$ without losing the service state, all replicas have access to a local stable storage device.
Any data successfully stored in such a device will not be lost in the advent of a recoverable crash fault.
Nonetheless, this guarantee does not extend to Byzantine faults, since a malicious replica is able to overwrite/corrupt its own stored data.


\section{Limitations of SMR as Blockchains}
\label{sec:limitations}

Blockchains and SMR share strong similarities since the main objective of both is to run a replicated deterministic service that executes transactions in total order.
However, even if we put aside consensus protocol properties, such as finality, commit latency, and scalability~\cite{abraham2017,Vukolic2016}, there are still important features blockchain applications need that SMR systems do not necessarily implement.
For example, blockchain applications need to maintain a self-verifiable persistent ledger with the executed transactions and support reconfigurations on the group of replicas, two features not present in most SMR implementations.

This section assesses the hindrances of the classic SMR model when supporting blockchain applications.
We start by presenting a ubiquitous digital coin application used in our evaluation.
Afterward, we analyze some experimental results that highlight the performance limitations of this blockchain application.




\subsection{SMaRtCoin}
\label{sec:smartcoin}

To demonstrate the inherent inefficiencies of SMR for supporting blockchain applications, we developed SMaRtCoin, a digital coin application on top of~\smart.
SMaRtCoin was broadly inspired by Bitcoin and more specifically by FabCoin. The latter being an application used to benchmark Hyperledger Fabric~\cite{Androulaki2018}. This application represents the simplest useful blockchain application we are aware of.

SMaRtCoin is a deterministic wallet-like service that manages coins based on the UTXO (Unspent Transaction Output) model introduced in Bitcoin~\cite{Nakamoto_bitcoin:a}.
In this model, each object (coin) represents a certain amount of currency possessed by a user. 
This means that a transaction consumes a given number of input objects to produce a number of output objects.
Therefore, this service supports two basic transaction types: $\texttt{MINT}$, used to create a certain amount of coins for a given address, and $\texttt{SPEND}$, to transfer coins to other addresses.
The state of the service is comprised of a table with the coins assigned to each address in memory and a list of addresses authorized to create new coins.

$\texttt{MINT}$ operations require the public key of the account that issued the transaction and the value of each coin to create for the issuer.
For that, the issuer needs to have permission to execute this operation, i.e., its public key must be in the list of authorized addresses to issue $\texttt{MINT}$ transactions which is defined in the genesis block.
$\texttt{SPEND}$ operations require the issuer's public key, the id of the coins that will be used as input and a set of key-value pairs each containing a public key of an account and the amount of coins it will receive.
Both types of requests need to be signed to ensure their authenticity and thus prove the ownership of the affected funds.


We implemented SMaRtCoin as a \smart service, using the $\mathit{invoke}$ and $\mathit{execute}$ interfaces provided by the library~\cite{Bessani2014}.
Clients generate signed SMaRtCoin transactions and submit them for the \smart ordering protocol.
This protocol runs a Byzantine consensus to order a batch of operations, instead of a single one.
Therefore, each replica receives a batch of transactions from the library's ordering protocol and delivers it to SMaRtCoin.
If SMaRtCoin successfully verifies that the client that issued the transaction has the right to execute it (e.g., it is the owner of the coins being transferred), the transaction is executed.

After transactions execution, a block containing the delivered batch together with the transactions results is created and appended to the ledger.
Once this block is synchronously written to stable storage, each replica replies to the clients with the results associated to each executed transaction.

\subsection{SMaRtCoin Limitations}

The experience of designing and running SMaRtCoin on top of \smart lead us to the observation of several gaps between the classic SMR and blockchain models.

\paragraph{Observation 1 (Performance issues)}
We run a set of experiments using different setups of SMaRtCoin on top of \smart.
Table~\ref{table:smartcoin-results} reports the throughput for SMaRtCoin when writing its blockchain synchronously and asynchronously to stable storage, considering different transaction signature verification strategies.
The experimental setup and methodology are detailed in Section~\ref{sec:evaluation}.
For these experiments, we configured the system with four replicas to tolerate a single Byzantine failure.

\begin{table}[!t]
\renewcommand*{\arraystretch}{1.3}
\centering
\caption{SMaRtCoin average throughput (txs/sec) with different signature verification and storage strategies.} 
\label{table:smartcoin-results}
\begin{tabular}{c|cc|ccc}
\toprule
\multirow{2}{*}{\textbf{Tx. type}} & 
\multicolumn{2}{c|}{\textbf{Seq. Sign. Verification}} & \multicolumn{3}{c}{\textbf{Parallel Sign. Verification}} \\ \cline{2-6}
   & \textbf{sync.} & \textbf{async.} & \textbf{sync.} & \textbf{async.} &\textbf{Dura-SMaRt} \\ \midrule
$\texttt{MINT}$  & \makecell{1801 \\ $\pm$ 321} & \makecell{1821 \\ $\pm$ 82} & \makecell{4079 \\ $\pm$ 152} & \makecell{4149 \\ $\pm$ 187} & \makecell{15015 \\ $\pm$ 422} \\ \midrule
$\texttt{SPEND}$ & \makecell{1729 \\ $\pm$ 302} & \makecell{1760 \\ $\pm$ 213} & \makecell{3881 \\ $\pm$ 177} & \makecell{4027 \\ $\pm$ 205} & \makecell{14829 \\ $\pm$ 549} \\ \bottomrule
\end{tabular}
\end{table}

In order to compare the results with other works, it is important to consider the size of the messages exchanged since this factor significantly affects the performance of BFT protocols~\cite{Bessani2014,Sousa2018}.
For $\texttt{MINT}$ operations, the requests and replies have an average size of 180 and 270 bytes, respectively.
For $\texttt{SPEND}$ operations, the size of the request is around 310 bytes, and the replies are 380 bytes long.
The size of the replies also approximates the space taken up by a serialization of each transaction (according to its type) in the ledger.

As can be seen on the left side of Table~\ref{table:smartcoin-results}, there is not much difference between the performance of the system with synchronous or asynchronous writes to stable storage when the signature of the coin objects is done sequentially, i.e., inside the state machine.
However, if we push this verification to the \smart message verification pool of threads~\cite{Bessani2014}, effectively exploiting the multiple cores of our servers to verify signatures in parallel, we improve throughput more than twice, moving the bottleneck to the blockchain stable storage.
We remark that signature verification can be further improved by parallelizing it through different replicas~\cite{Crain18b}.

Although parallel signature verification significantly improves system performance, if we remove the blockchain durability implementation out of the SMR application layer, and instead use the \smart durability layer~\cite{Bessani13}, we still have similar guarantees in terms of service durability, but the performance improves more than $3.6\times$.
As explained in Section~\ref{sec:bftsmart-statetransfer}, this gain is due to the fact that the \smart durability layer accumulates several batches of transactions before delivering them to the SMR service for processing while writing these batches in a single IO operation.


\paragraph{Observation 2 (SMaRtCoin does not implement an immutable ledger)}
It is worth pointing out that, in all the scenarios evaluated so far, there is no immutable ledger that could be fetched to verify transactions. 
%
This happens because writing synchronously to stable storage only during the execution of the state machine and before sending a reply to the client, ensures only what we call \emph{external durability}:
an executed operation is never reversed after the client see its completion~\cite{Bessani13}.
In other words, an operation is durable only if the client that issued it receives matching replies from a ($f$-dissemination) Byzantine quorum with $\lfloor \frac{cv.n+cv.f+1}{2} \rfloor \geq 2cv.f+1$ replicas~\cite{Mal98}.
This ensures that these replicas wrote the operation in their logs and, even if there is a full crash and recover of the system, any other Byzantine quorum will see this operation on the log of at least one correct replica and recover the state with such operation.
Notice a single log is enough because each value decided in \smart comes with a proof that it was the result of a consensus, as discussed in Section~\ref{sec:bftsmart}. 
The consequence of this guarantee is that a single durable log of a replica does not provide a \emph{durable committed} history of the system execution, as a suffix of the logged operations can be undone.
To be sure some logged operation will not be undone, one needs to check logs from a Byzantine quorum of replicas.
What is missing here is \emph{log self-verifiability}, i.e., verifying a single correct log should be enough for obtaining the complete execution on history of the system up to that point.

\paragraph{Observation 3 (Reconfiguration depends on a centralized authority)}
Most BFT SMR systems assume a static set of nodes participating in the ordering protocol~\cite{Cas99, Abd05,Cow06,Kot07,Amir2011,Ver09,Veronese13,Aublin2013,Behl2017}.
However, this is not suitable for a blockchain platform, since the set of nodes participating in the consortium are expected to change during the lifespan of the system.
Moreover, there are indeed a few SMR systems that are prepared to accept new replicas to join the system and older ones to leave it, but they rely on a centralized third party with administrative privileges~\cite{Lorch2006,Rodrigues2012,Ongaro2014}.
This is also not well suited for blockchains, since nodes should have the ability to join and leave in an autonomous way.

\section{\smartchain}
\label{sec:smartchain}

\smartchain is a blockchain platform based on \smart that efficiently support applications such as the digital coin described in Section~\ref{sec:smartcoin}. 
\smartchain addresses the aspects discussed in the previous section, with two novel mechanisms: the blockchain storage layer, and the decentralized reconfiguration protocol.
Before diving into the details about them, we present an overview of what need to be done to transform SMR to blockchains.

\subsection{Overview: Transforming SMR to Blockchains}
\label{sec:inefficiencies}

The previous limitations show that naively implementing a blockchain application, even the simplest one, can result in a low-performance system with some missing features, independently of how good is the consensus being employed.
Observation 1 shows that beside the scalability issues~\cite{Vukolic2016,Guerraoui2019}, which have been the main focus of most of the recent work on BFT replication, it is also important to ensure that the system (1) can deal efficiently with messages of significant size, (2) is able to exploit multi-cores for cryptographic operations, and (3) implements an effective durability layer.
Observations 2 and 3 are more complex to overcome and require addressing two fundamental issues on state-of-the-art SMR systems.


\subsubsection{Turning Operation Logs into Blockchains}

Practical SMR systems require the usage of an internal log of delivered requests, both to recover from a faulty leader and to enable the transference of service state to recovered replicas~\cite{Cas99,Bessani2014}.
Three requirements must be addressed to transform such internal log into a blockchain.

Firstly, this log must be durable.
It is necessary to carefully devise a solution for log durability in order to ensure that synchronous writes to disk do not cripple system performance~\cite{Bessani13}.
Furthermore, to approach the idea of blocks, logs should no longer be comprised of individual operations, and instead composed by a sequence of blocks with the transactions ordered by the underlying protocol.
Most existing SMR protocols already assume that batches of transactions are ordered on each consensus, thus making the notion of blocks quite natural.
In addition, each entry in the log will require a block header and a certificate that renders the block/log self-verifiable.
Moreover, request processing and block persistence must be decoupled to ensure log self-verifiability (as defined before) and not only \smart external durability.

The second requirement is related to state snapshots. 
Most systems truncate the log when snapshots are created.
In a blockchain platform, snapshots would allow a fast (re)initialization of replicas.
Thus, the file in which they are stored should be linked with the chain of blocks. 

Finally, the result of the transaction execution must also be stored within each block to enable auditability of transactions, matching the blockchain model.


\subsubsection{Reconfiguring the Set of Nodes}

As discussed before, most BFT SMR systems assume a static set of replicas, and the few that are prepared to accept replica group changes rely on a centralized third party with administrative privileges~\cite{Bessani2014,Rodrigues2012}.
Such centralized management goes against the distributed trust promised by blockchains.
A more appropriate solution for a blockchain scenario would be to enable the nodes themselves to judge if another node can join the system.
In addition, this mechanism should be designed in such a way that the criteria by which nodes are allowed to join should be specified by the blockchain application.

An additional problem associated with reconfigurations is how to ensure the security and verifiability of the blockchain data structure when the set of keys that validate blocks change.
More specifically, new mechanisms must be designed to impede (malicious) nodes removed from the consortium to create forks on the blockchain.

\subsection{The Blockchain Layer}

This section details how the issues previously discussed can be addressed in a blockchain design.
We start by defining the blockchain data structure and then we proceed with an in-depth discussion on how it can be extended with new transactions, checkpoints, and consortium changes.

\begin{figure}[t]
\centering
\includegraphics[scale=1.0]{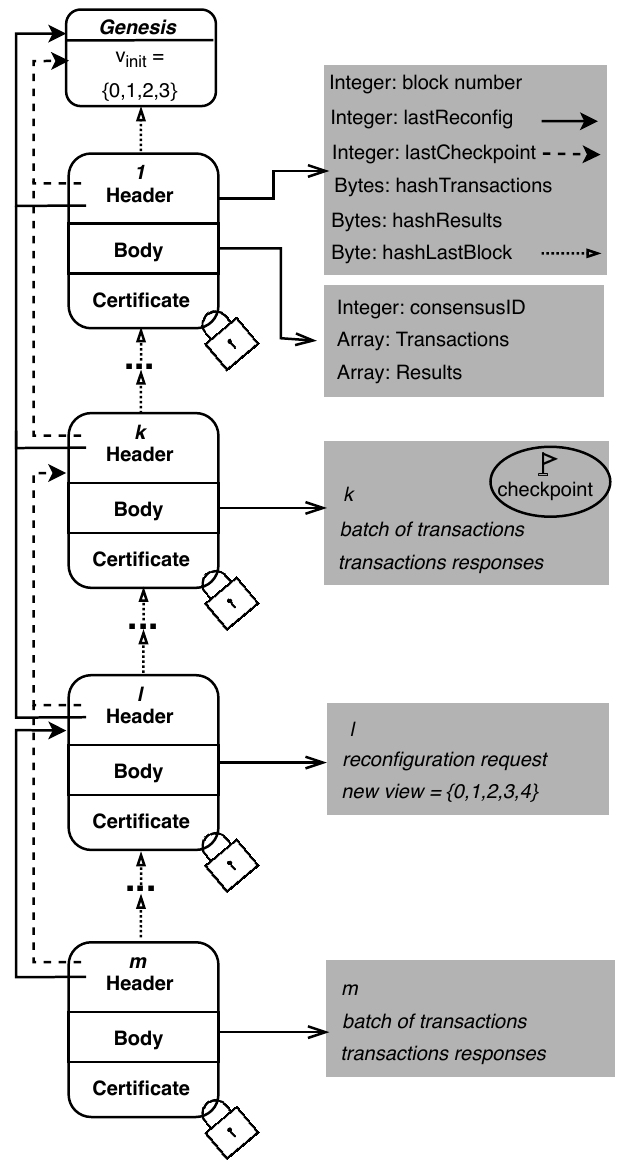}
\caption{\smartchain blockchain structure.}
\label{fig:block-structure}
\end{figure}

\subsubsection{Blockchain structure} 

Figure \ref{fig:block-structure} illustrates the structure of the blockchain maintained by \smartchain.
On the top of the figure (block $1$) we have a detailed description of a block, which is composed of three parts: (1) a header containing block metadata, (2) a body containing the list of transactions decided in a consensus instance and associated results, and (3) a certificate with a cryptographic proof of the block validity.

The header is composed of three integers representing the block number, the number of the block containing the last reconfiguration, and the number of the block in which the last service snapshot took place.
Moreover, the header also contains hashes of the batch of transactions in the block body, the results of the execution of these transactions, and the previous block.

The body of the block contains the metadata of the consensus that delivered a batch of transactions (e.g., the consensus instance number), the list of transactions on this batch, and the list of results of each one of these transactions.\footnote{Results can include a compact representation (e.g., a Merkle tree) of the state changes caused by the transactions, making SMaRtChain compatible with execution engines like the Ethereum Virtual Machine, as in SBFT~\cite{Golan-Gueta2019}.}

The certificate comprises a set of $\lfloor \frac{cv.n+cv.f+1}{2} \rfloor \geq 2cv.f+1$ signatures of the block header generated by different replicas in the current view.
In a SMR-based blockchain system this certificate suffices to guarantee that there is no other block that can be generated in this position on the blockchain.

\subsubsection{Extending the Blockchain}

The system starts with a genesis block containing the initial members of the consortium, their public keys, and other setup data.
Every time a batch of transactions is delivered in total order and executed by the blockchain application, a new block is created containing the batch itself and the results of each transaction.
This can be seen in blocks $1$, $k$, and $m$ in Figure~\ref{fig:block-structure}.

\subsubsection{State Checkpoints}

In order to accelerate the launching of new consortium members or decrease the time to repair crashed replicas, \smartchain employs durable checkpoints, stored outside the blockchain.
A checkpoint contains a snapshot of the application state and a reference to the \emph{last block covered by it} (block $k$ in Figure~\ref{fig:block-structure}), i.e., the most recent block whose transactions were executed before the snapshot was taken.
This means that a checkpoint makes the blocks before it mostly obsoletes for starting a replica.

\smartchain requires a checkpoint to be created after a sequence of $z$ blocks are processed.
The parameter $z$ is defined in the genesis block.
This is different from traditional SMR systems, in which the checkpoint is defined based on the number of transactions executed. 
We changed it to blocks to avoid having checkpoints that partially cover a block.

Each block $b$ stores the number $c$ of the last block for which its transactions were included in the most recent checkpoint at the time $b$ was created.
This is important to inform anyone reading the blockchain that there is a state snapshot that represents the state of the system up to block $c$ (inclusive).

\subsubsection{Consortium Changes}

A fundamental characteristic of permissioned blockchains is that members of the consortium know each other. 
A simple way to do that is by storing the current composition of the consortium on the blockchain.

Our blockchain structure accommodates that in two ways.
First, by storing the initial consortium composition in the genesis block.
Second, by storing the transaction that reconfigures the system and the corresponding new view, in a separated \emph{reconfiguration block} (see block $l$ in
Figure~\ref{fig:block-structure}).
Similarly to the checkpoint approach, each block stores the number of the last reconfiguration block before it in the chain.
This ensures blockchain verifiers have access to enough public keys that validate the certificate of each block created in the view.

\subsection{Strengthening the Blockchain Persistence}

As discussed before, \smart provides only external durability, i.e., a transaction is irreversibly committed only if its issuer sees matching replies from a quorum of replicas (see Observation 2 in Section~\ref{sec:limitations}).
This limitation also affects our blockchain architecture if no changes are made.

Considering the definition of blockchain in terms of Persistence and Liveness (Section~\ref{sec:blockchain}), this external durability is equivalent to $1$-Persistence, i.e., only the second to last block is immutable.
However, there are other possibilities:

\begin{itemize}

\item \textit{$0$-Persistence}: Perfect durability, once a block is written, it is immutable.

\item \textit{$\alpha$-Persistence}: Standard durability, with $\alpha$ being the number of consensus instances running in parallel in the system.
\smart runs a single consensus at time ($\alpha = 1$), as described before.

\item \textit{$\lambda$-Persistence}: Durability provided when using asynchronous stable storage writes.
The value of $\lambda$ is dependent on the environment but clearly a small integer greater than zero.

\item \textit{$6$-Persistence}: The durability provided (with high probability) in the Bitcoin's blockchain~\cite{Nakamoto_bitcoin:a}.

\item \textit{$\infty$-Persistence}: No durability, provided when storing blocks only in memory.

\end{itemize}

In this paper we are particularly interested in achieving \textit{$0$-Persistence}, a guarantee similar to the durability provided by most database systems.
To do that, we need an additional communication step on the system, just after the transactions are executed and persisted.
This extra round of communication -- designated as $\texttt{PERSIST}$ phase -- consists in making each replica generate its own signature of the block (which will now include the aforementioned transaction results) and disseminate these signatures among the view.
Once a replica collects $\lfloor \frac{cv.n+cv.f+1}{2}\rfloor$ signatures for the same block, it appends these signatures to the block, thus creating a certificate for it.
Notice that this write is asynchronous since if all replicas crash after synchronously writing the header and body of a block, when they recover the only possible next action is to create the same certificate again.

\begin{figure}[!t]
	\centering
    \includegraphics[page=11,scale=0.8]{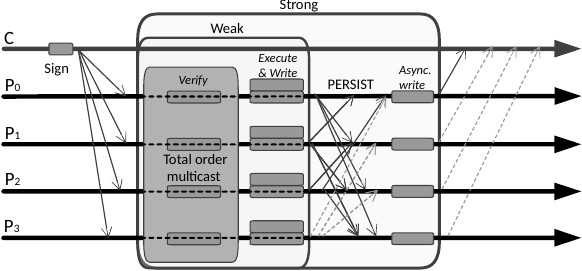}
    \caption{\smartchain message pattern.}
    \label{fig:smartchain-pattern}
\end{figure}

This modification ensures $0$-Persistence because the block is considered written only when a replica knows that a Byzantine quorum of replicas executed and recorded the same set of transactions to their stable storage.
Consequently, even if all replicas crash and recover, these transactions will still be visible in the blockchain.

\smartchain supports either $0$- or $1$-Persistence, in variants we call \textit{weak} and \textit{strong}, respectively.
Figure \ref{fig:smartchain-pattern} illustrates the message pattern of both variants.
For both cases, the algorithm for state transfer is basically the same as used in \smart (Section~\ref{sec:bftsmart-statetransfer}), sending the last checkpoint covering up to a block $b$ plus the blocks after it.

\subsection{The Reconfiguration Protocol}

\smartchain provides a new reconfiguration protocol that does not rely on a trusted third party to manage reconfigurations, allowing replicas to join/leave the system in an autonomous and secure way, following application-specific conditions.

\begin{figure}[t]
	\centering
    \includegraphics[scale=0.7]{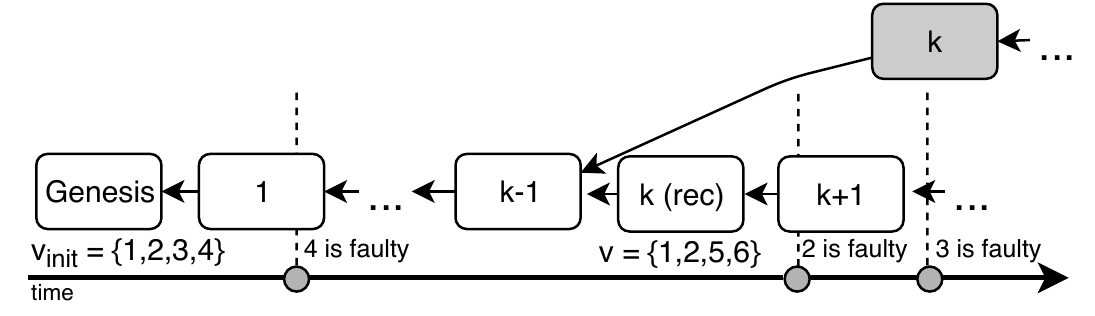}
    \caption{Fork created by malicious processes.}
    \label{fig:fork}
\end{figure}

An important aspect related with reconfigurations is how to avoid forks caused by faulty nodes removed from the system.
Recall that our assumption is that in each active view $v$, there is at most $v.f$ faulty nodes.
However, we do not assume anything about the nodes from past views.
Figure~\ref{fig:fork} shows an example where the failure thresholds of all views are respected, but in which node $3$, that is compromised after being removed from the system, together with faulty nodes $2$ and $4$ (also removed), is able to create a fork after block $k-1$ by extending the blockchain without the reconfiguration block $k$.

In \smartchain, we solve this problem by decoupling replicas permanent key pairs from their \emph{consensus key pairs}, which are used to create a consensus decision proof and also to obtain a block certificate.
The idea is to make all replicas generate new consensus key pairs for each view they participate, certifying each generated public consensus key with their permanent private keys, and discard their consensus key pairs on each view change.
This forgetting protocol~\cite{Mar04} ensures that even if a replica becomes faulty later, after a new view is installed, it cannot recover the discarded consensus private key and thus cannot vouch for a block in some old view (as done by nodes $3$ and $4$ in the example).

The consensus public keys for a new view need to be stored in the reconfiguration block, together with the list of members of the view.
This requires the inclusion of these keys in the reconfiguration transaction.
However, to preserve reconfiguration liveness in non-synchronous systems, the processes handling the reconfiguration transaction(s) that will install a new view $v$ are ensured to collect at most $v.n-v.f$ of such keys.
Fortunately, this quorum is enough for avoiding forks since, in the worst case, it will contains $v.f$ keys from faulty processes and a collusion with the $v.f$ processes whose keys were not included in the reconfiguration block (that can become malicious later) will not be enough to generate a valid proof for a consensus decision or to certify a block, which requires $\lfloor \frac{cv.n+cv.f+1}{2} \rfloor$ signatures.
It is worth to mention that correct processes whose keys are not included in the reconfiguration block but that participate in the view need also to forget old keys and generate new ones.
These new keys are disseminated in the first messages these processes send in the new view.

\begin{figure}[t]
	\begin{center}
    \begin{subfigure}{0.4\textwidth}
        \includegraphics[page=7,scale=0.72]{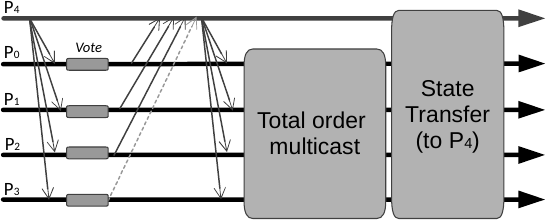}
        \caption{Join message pattern.}
        \label{fig:join-pattern}
    \end{subfigure}
    \begin{subfigure}{0.4\textwidth}
        \includegraphics[page=8,scale=0.72]{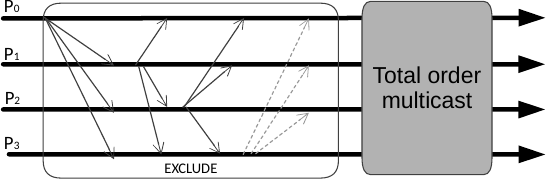}
        \caption{Exclusion message pattern.}
        \label{fig:exclusion-pattern}
    \end{subfigure}
    \end{center}
    \caption{\smartchain reconfiguration protocol.}
\end{figure}

Concretely, for a new node to join the system the following steps need to be executed (Figure~\ref{fig:join-pattern}): (1) it asks the nodes in $cv$ for a permission to join the system; (2) each node may accept or not the request based on an application-specific policy (e.g., the new node is certified by a trusted third party, it solved a proof-of-work, or it acquired a certain amount of the blockchain-specific cryptocurrency), by sending a signed reply to the joining node which also contains its new public key to be used in the next view; (3) if the joining node receives signed acceptance replies from a quorum of $cv.n-cv.f$ nodes in $cv$, it assembles a certificate and invokes a reconfiguration transaction that goes through the ordering protocols.
After this join transaction is executed and the new node is included in the current view, its state is updated as previously described.

If a node decides to leave the system by itself, it collects public keys for a new view without itself from a quorum of nodes and notifies the others by submitting a special leave transaction in total order.
Once a node receives this transaction, it generates a new view with that node excluded from the group.
On the other hand, if the group decides to remove some node from the system, each node submits a special remove transaction to the ordering protocol asking for that exclusion and informing its public key for the new view (Figure~\ref{fig:exclusion-pattern}).
Once a node observes $cv.n-cv.f$ of such transactions from different nodes targeting the same node, it generates a new view without that node.
Notice that the overhead of requiring all these transactions for running a single reconfiguration will be limited due to batching.

\subsection{Consolidated Algorithm} 

Algorithm~\ref{alg:smartchain} consolidates all the previous ideas into a single module to be run on top of the consensus layer. 
During initialization, several variables are initialized and the genesis block with all consensus public keys of the initial view is written to stable storage (lines 1-10).
Every time the ordering protocol delivers a batch of transactions, they are stored together with the respective consensus proofs (see Section~\ref{sec:bftsmart}) to disk (line 18).
The \texttt{asyncWriteBC} command denotes the action of asynchronously writing data to the blockchain stored in disk.
Moreover, the transactions are delivered to the application code for execution (line 19) and the results are also stored to disk (line 20).
This effectively creates the block's body.
Since writing transactions to disk is done before executing them, asynchronously, storage and execution are performed in parallel.
Finally, the header is written to close the block (lines~21, 26-29), the replies are sent to the clients (lines 22-23), it is verified if a snapshot of the service must be created (line 24), and the blockchain becomes ready to receive the next block (line 25).

Additionally, in the strong variant, the block certificate is also created and stored in the block (lines 31-34). 
More specifically, each replica sends a signed $\texttt{PERSIST}$ message
with the hash of the block header to all replicas in $cv$.
Once a replica receives correctly signed $\texttt{PERSIST}$ messages from a quorum of replicas in $cv$, it creates a certificate that authenticates the block and writes it to disk. 


\begin{algorithm}[!t]
\caption{\smartchain Algorithms}
\label{alg:smartchain}

\SetAlgoVlined

\SetKwFunction{send}{send}
\SetKwFunction{hash}{hash}
\SetKwFunction{deliver}{deliver}
\SetKwFunction{todeliver}{totalOrderDeliver}
\SetKwFunction{execute}{execute}
\SetKwFunction{enqwrite}{asyncWriteBC}
\SetKwFunction{enqwriteSN}{asyncWriteSN}
\SetKwFunction{closeblock}{closeBlock}
\SetKwFunction{syncdisk}{syncDisk()}
\SetKwFunction{takesnap}{takeSnapshot()}
\SetKwFunction{resetCached}{resetCached()}
\SetKwFunction{writeGen}{writeGenesisBlock()}
\SetKwFunction{ckpt}{checkpoint()}
\SetKwFunction{wait}{wait until}

\SetKwIF{If}{ElseIf}{Else}{if}{}{else if}{else}{end if}

\SetKwFor{upon}{Upon}{do}{end}
\SetKwFor{proc}{Procedure}{}{end}

\newcommand\mycommfont[1]{\footnotesize{#1}}
\SetCommentSty{mycommfont}

\DontPrintSemicolon
\scriptsize 

\upon{Init}{

    $\mathit{myId} \leftarrow$ replica identifier \tcp*{replica identifier}
	$\mathit{bNum} \leftarrow 1$  \tcp*{next block number}
	$\mathit{lRec} \leftarrow -1$ \tcp*{last reconfiguration block number}
	$\mathit{lCkp} \leftarrow -1$ \tcp*{last checkpoint block number}
	$\mathit{lbHash} \leftarrow \hash(\emptyset)$ \tcp*{hash of the last block}
	$\mathit{lSnapshot} \leftarrow \perp$ \tcp*{last state snapshot taken}
	$\mathit{cv} \leftarrow v_{init}$ \tcp*{system current view}
	\resetCached \tcp*{resets the cached data}
	\writeGen \tcp*{writes the genesis block to disk}
}

\BlankLine

\proc{\resetCached}{

    $\forall i \in N : \mathit{Txs}[i] \leftarrow \emptyset$ \tcp*{transactions for each block i}
	$\forall i \in N : \mathit{Res}[i] \leftarrow \emptyset$ \tcp*{responses for transactions on each block i}
	$\forall i \in N : \mathit{Cert}[i] \leftarrow \emptyset$ \tcp*{certificates for each block i}
	$\forall i \in N : \mathit{Headers}[i] \leftarrow \emptyset$ \tcp*{headers for each block i}
}

\BlankLine

\upon{\todeliver $\langle\mathrm{BATCH}, \mathit{cid}, \mathit{txs}[], \mathit{proofs}[] \rangle$} {
    
    
    $\mathit{Txs[bNum]} \leftarrow \langle \mathit{txs[], proofs[]} \rangle$\;
    
    $\enqwrite(\langle \mathit{cid, Txs[bNum]} \rangle)$\;
    $\mathit{Res[bNum]} \leftarrow \execute(\mathit{Txs[bNum]})$\;
    $\enqwrite(\mathit{Res[bNum]})$\;
	
    $\closeblock(\langle \hash(\mathit{Txs[bNum]}),\hash(\mathit{Res[bNum]})\rangle)$\;
    
    \ForEach{$\langle \mathit{clientId, res} \rangle \in \mathit{Res[bNum]}$}{
        \send $\langle \mathrm{REPLY}, \mathit{res} \rangle$ to $\mathit{clientId}$\;
    }
   
    $\ckpt$ \; 
    $\mathit{bNum++}$\;
}

\BlankLine

\proc{$\closeblock(htx,hres)$}{

	$\mathit{Headers[bNum]} \leftarrow \langle \mathit{bNum, lRec, lCkp, htx, hres, lbHash} \rangle$\;

	$\enqwrite(\mathit{Headers[bNum]})$\;
	 
	\syncdisk \;
	
	$\mathit{lbHash} \leftarrow \hash(\mathit{Headers[bNum]})$\;
	\uIf{$\mathtt{STRONG}$ $\mathtt{PERSISTENCE}$}{
	   \send $\langle \mathrm{PERSIST}, \mathit{bNum, \langle myId,lbHash \rangle_{\sigma_\mathit{myId}}}\rangle$ to $cv$\;
	   \textbf{wait until} $|\mathit{Cert[bNum]}| \geq \lceil\frac{\mathit{cv.n+cv.f+1}}{2}\rceil$ \;
	   \enqwrite($\mathit{Cert[bNum]}$)\;
	}
}

\BlankLine

\upon{\deliver $\langle \mathrm{PERSIST}, \mathit{bNum, \langle r,lbHash \rangle_{\sigma_r}}\rangle$} {

    $\mathit{Cert[bNum]} \leftarrow \mathit{Cert[bNum]} \cup \{\langle \mathit{r,lbHash} \rangle_{\sigma_r}$\}\;

}

\BlankLine

\upon{\todeliver $\langle \mathrm{VIEW}, \mathit{cid, recTx, recProof, nKeys[]} \rangle$} {
    \uIf{$\mathit{valid(recTx,recProof,nKeysx[])}$}{
        $\mathit{Txs[bNum]} \leftarrow \langle \mathit{recTx, recProof, nKeys[]} \rangle$\;
        $\enqwrite(\langle \mathit{cid, Txs[bNum], nKeys[]} \rangle)$\;
        updates $\mathit{cv}$ according to $\mathit{recTx}$\;
        $\mathit{Res[bNum]} \leftarrow \langle \mathit{recTx.senderId, cv} \rangle$\;
        $\enqwrite(\mathit{Res[bNum]})$\;
    
	    \closeblock($\hash(\mathit{Txs[bNum]}),\hash(\mathit{Res[bNum]})$)\;
	    \send $\langle \mathrm{REPLY}, \mathit{cv} \rangle$ to $\mathit{recTx.senderId}$\;
        $\mathit{lRec} \leftarrow \mathit{bNum}$\;
        $\ckpt$ \;
        $\mathit{bNum++}$\;
    }
}

\BlankLine

\proc{\ckpt}{
    \uIf{$(\mathit{bNum}~\%~\mathrm{CHECKPOINT\_PERIOD})~=~0$} {
        $\mathit{lCkp} \leftarrow \mathit{bNum}$\;
        \resetCached\;
        $\mathit{lSnapshot} \leftarrow \takesnap$\;
        $\enqwriteSN(\mathit{lSnapshot})$\;
    }
}

\BlankLine

\upon{\deliver $\langle \mathrm{ST\_REQ}, \mathit{cid, stateReq} \rangle$} {
    $\mathit{lastTxs} \leftarrow$ get transactions from $\mathit{lCkp}+1$ to $\mathit{cid}$ from the cache\;
    \send $\langle \mathrm{ST\_REP}, \mathit{cid, lastTxs, lSnapshot} \rangle$ to $\mathit{stateReq.senderId}$\;
}
\end{algorithm}

Membership updates are stored in their own blocks (lines 37-48). The algorithm presents the processing needed to include or remove a process that asked to join or leave the system, respectively.
The processing to exclude a member from the system is similar, but in this case, it is necessary to wait for transactions from a quorum of nodes advocating for the removal.

Finally, snapshots are written outside the blockchain in a different file (line 54) and state transfer requests are replied with the last snapshot together with the blockchain data cached since the last checkpoint (lines 55-57).

\section{Evaluation}
\label{sec:evaluation}

We implemented \smartchain over \smart and conducted several experiments \textit{(1)}~to compare the performance of different strategies for blockchain data persistence, \textit{(2)}~to compare the \smartchain performance with similar systems (Tendermint and Hyperledger Fabric), and \textit{(3)}~to understand the system behavior under events like reconfigurations, crashes, and recoveries.

\subsection{Experimental Setup and Methodology} 

The experimental environment was configured with 14 machines connected to a 1Gbps switched network.
The machines were configured with Ubuntu Linux 16.04.5 LTS operating system and JRE 1.8.0, hosted in Dell PowerEdge R410 servers.
Each machine has 32 GB of memory and two quadcore 2.27 GHz Intel Xeon E5520 processor with hyperthreading, i.e., supporting 16 hardware threads. 
The machines have also a 146 GB SCSI HDD (Seagate Cheetah ST3146356SS). 
The experiments were conducted in up to ten replicas hosted in separate physical machines.
Moreover, 2400 client processes were distributed uniformly across the other four machines.



\smartchain was configured to use a maximum batch size (block size) of 512 transactions.
The experiments were conducted in two phases: the first one is composed of $\texttt{MINT}$ operations to generate new coins, and then a second phase considers $\texttt{SPEND}$ operations to transfer the generated coins to new addresses.
Following the UTXO model, this corresponds to single-input, single-output $\texttt{SPEND}$ transactions.
Each client issued up to 1000 requests of each type ($\texttt{MINT}$ and $\texttt{SPEND}$).
In this section we report only the values for $\texttt{SPEND}$ since both types of transactions yield equivalent results.

For each experiment, the throughput was measured at the replicas at regular intervals (at each 10k operations).
From the collected data, 20\% of the values with greater variance were discarded and the average values are presented.
Standard deviations were always under 500 txs/sec.

\begin{figure*}[t]
	\centering
    \begin{subfigure}{0.325\textwidth}
        \includegraphics[scale=0.62]{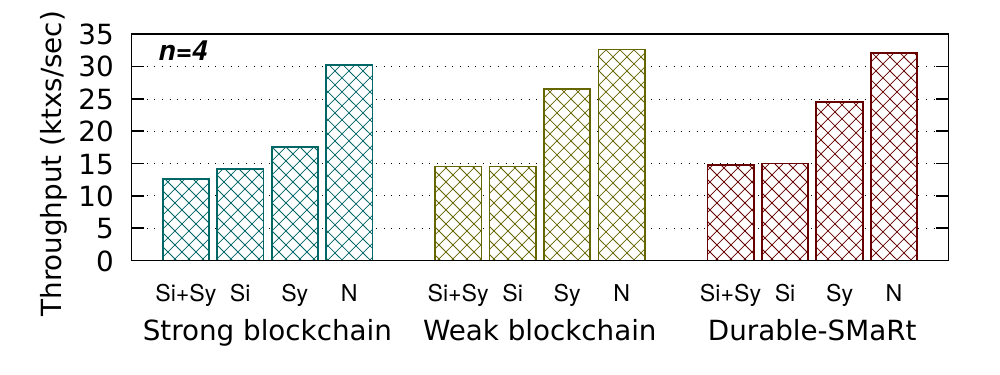}
        \label{fig:throughputF1}
    \end{subfigure}
    \begin{subfigure}{0.325\textwidth}
        \includegraphics[scale=0.62]{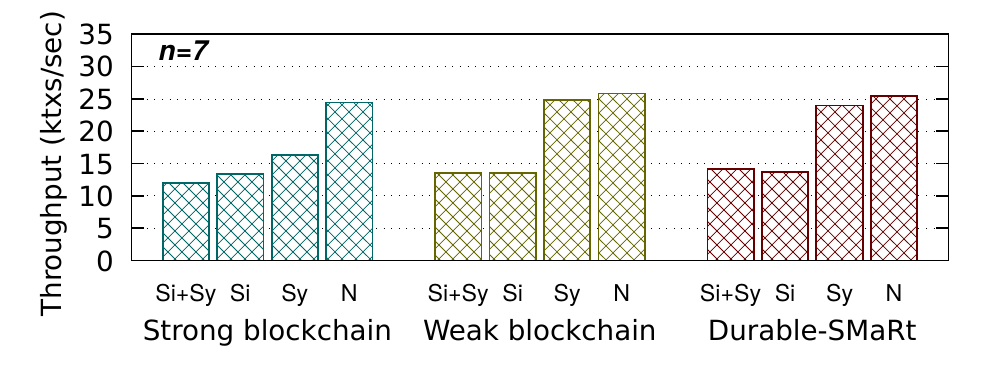}
        \label{fig:throughputF2}
    \end{subfigure}
    \begin{subfigure}{0.325\textwidth}
        \includegraphics[scale=0.62]{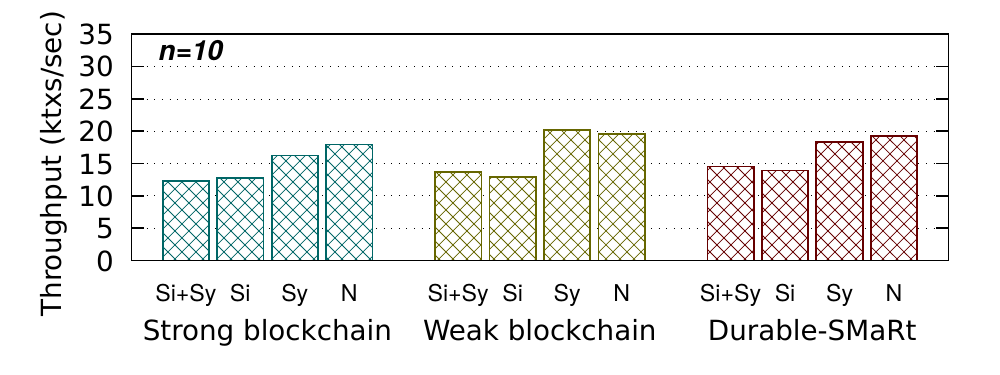}
        \label{fig:throughputF3}
    \end{subfigure}
    \caption{\smartchain throughput for different consortium sizes and blockchain persistence guarantees. Legend: Si+Sy = Signatures and synchronous writes; Si = Signatures only; Sy = Synchronous writes only; N = None.}
\label{fig:throughput}
\end{figure*}

\subsection{Results}

This section presents the experimental results, which were divided in three subsets according to the evaluation goals.

\paragraph{Comparing different blockchain strategies}
We compared the system performance considering different blockchain persistence guarantees: \smartchain configured with synchronous storage writes ($0$- and $1$-Persistence in the strong and weak variants, respectively), asynchronous storage writes ($\lambda$-Persistence for both variants), and memory only ($\infty$-Persistence for both variants).
As a baseline, we also present results for the efficient durability layer of \smart~\cite{Bessani13}, which does not implement a blockchain (Section~\ref{sec:smartcoin}).
Figure~\ref{fig:throughput} presents the throughput results for all these configurations considering different consortium sizes and the use or not of signatures.

The results show that signature verification represents the major factor that impacts performance, followed by the storage strategy.
For $n=4$ and when using signatures, \smartchain throughput reaches around 12k and 14k txs/sec for the strong and weak variants, respectively.
When signatures are disabled, these values increase to around 18k and 26k txs/sec in the strong and weak variants, respectively. 
Notice that the size of transactions makes the throughput of plain \smart (N setup) reach 33k txs/sec, which is much less than the 80k txs/sec the system achieve with transactions of few bytes~\cite{Bessani2014}.

In our experiments, the size of the consortium has a minor impact on the performance of the configurations with stronger guarantees (signatures and synchronous writes), in all durability strategies.
This shows that the consensus protocol was not the bottleneck in these scenarios.
Instead, the bottleneck is the time demanded to write the ledger to disk and to perform signature verification. 
However, it is expected that the lack of scalability of \smart consensus protocol will make it a bottleneck in larger groups~\cite{Guerraoui2019}.


Likewise, the results show that the additional $\texttt{PERSIST}$ phase in the strong blockchain variant does not significantly impact system performance, as the obtained results for this setup are only $13\%$ lower than the ones obtained for the weak variant.

\paragraph{Comparison with other systems}
Table~\ref{table:comp_others} compares the \smartchain performance with two other 
well-known BFT blockchain systems: Tendermint~\cite{Buchman2016,Yackolley2018,Ethan2018} and Hyperledger Fabric~\cite{Androulaki2018} configured with a BFT ordering service~\cite{Sousa2018}.
For both variants, \smartchain was configured to use signatures and synchronous writes.
Both Tendermint and Hyperledger Fabric were also configured for maximum durability.
Finally, all systems were configured with four replicas to tolerate a single Byzantine failure.

Table~\ref{table:comp_others} shows that \smartchain performs significantly better than the competing systems.
Tendermint uses an architecture that decouples application and ordering layers, similar to SMaRtCoin, and the performance results were also similar (Section~\ref{sec:smartcoin}).
Although other works reported higher throughput for Hyperledger Fabric (e.g., approximately 1k txs/sec~\cite{Rusch2019}), we could reach at most 381 txs/sec in our testbed.

\begin{table}[!t]
\renewcommand*{\arraystretch}{1.3}
\centering
\caption{Throughput and latency for different blockchains.}
\label{table:comp_others}
\begin{tabular}{l|rr}
\toprule
   \textbf{Blockchain} & \textbf{Throughput (txs/sec)} & \textbf{Latency (sec)}  \\ \midrule
   \smartchain Strong  & 12560 $\pm$ 480 & 0.210 $\pm$ 0.033 \\
   \smartchain Weak  & 14547 $\pm$ 465 & 0.200 $\pm$ 0.023\\
   Tendermint  & 1602 $\pm$ 395 & 1.378 $\pm$ 0.421 \\
   Hyperledger Fabric  & 381 $\pm$ 102 & 1.602 $\pm$ 0.504  \\
\bottomrule
\end{tabular}
\end{table}

\begin{figure*}[t]
	\centering
    \includegraphics[scale=0.68]{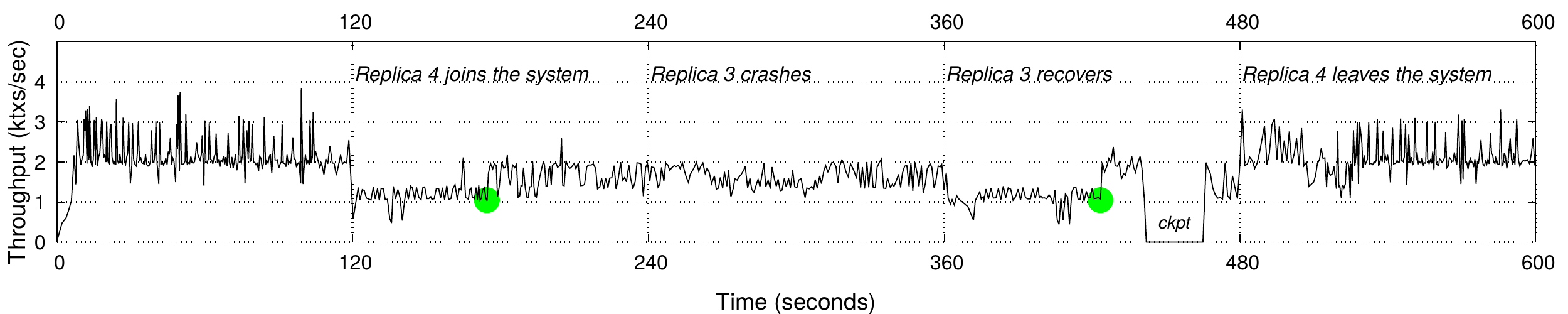}
    \caption{Throughput evolution across time and events, $v_{init} = \{0,1,2,3\}$. 
    }
    \label{fig:run}
\end{figure*}

\paragraph{Reconfigurations, crashes, and recoveries}
Figure~\ref{fig:run} shows the behavior of the strong variant of \smartchain, using signatures and synchronous writes, in a run with different events and 600 clients accessing the system.
For this experiment, the system was configured with 8 million UTXOs representing 10\% of the current number of UTXOs in the Bitcoin network~\cite{bc2019}, leading to a state of 1GB.

We can observe that the throughput increases until all clients become operational, around second 7.
At second 120, replica $4$ joins the system and the throughput decreases since large quorums are used in the protocol.
At second 240, replica $3$ crashes, which does not impact throughput, and later recovers at second 360.
In second 442, replicas perform a checkpoint that takes 23 seconds to finish.
During this period, the throughput drops to almost zero.
It is possible to configure replicas to take checkpoints at different instants in the execution to decrease its impact in the overall system performance~\cite{Bessani13}.
Finally, at second 480, replica $4$ leaves the system and throughput goes back to what was observed in the beginning of the experiment.


 \begin{figure}
 	 \centering
     \includegraphics[scale=1.4]{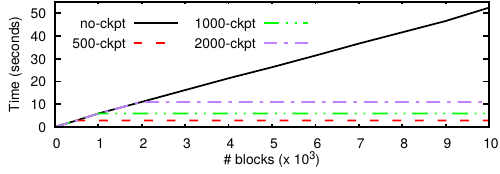}
     \caption{Time demanded to update a replica.}
     \label{fig:blocks}
 \end{figure}

Notice that after a join or a recovery, replicas demand approximately 60 seconds to obtain and install the 1GB state from the other replicas (green spots in Figure~\ref{fig:run}).
Throughput is slightly smaller during this period since replicas must send their state to the joining/recovering replica.
By using checkpoints and state transfer, a replica can join the system faster than in other systems that do not employ this technique.
For example, currently a node must process a blockchain of 223GB (9080186 blocks) to join the Ethereum network~\cite{et2019}, even pruning old states. 
Based on this observation, Figure~\ref{fig:blocks} shows the processing time demanded to update a replica considering different checkpoint periods and blockchain sizes.
Checkpoints boost the reconfiguration performance since joining nodes need to process only the transactions logged after the last checkpoint.

\section{Related Work}
\label{sec:rel-work}


Since Bitcoin's inception and widespread adoption there have been an impressive amount of work on both permissionless and permissioned blockchain platforms.
Most of these works focus on the multiple types of blockchain consensus, but very few provide an in-depth discussion about blockchain durability and the issues with decentralized consortium reconfiguration.

\paragraph{Durability}
The scale, latency, and probabilistic finality of the most popular blockchains lead to an ad-hoc implementation of blockchain durability.
However, the recent popularization of small-scale permissioned blockchains (e.g.,~\cite{Buchman2016,Martino2016,Voell2016,chain2017,Androulaki2018}) and their use as distributed transaction platforms~\cite{Nathan19,Sharma19}, calls for a better understanding of blockchain durability.
However, to the best of our knowledge, this subject was not yet explored in both academic and industrial works.

One of the best known blockchain platforms is Hyperledger Fabric~\cite{Androulaki2018}.
The platform is designed to support pluggable implementations of different components, such as the ordering and membership services.
Fabric's key innovation is the execution of transactions before establishing a total order among (blocks of) them.
Only after such order is established the blocks are validated by the peers and then written to stable storage.
Fabric durability guarantees are not well documented, but the lack of coordination between peers during blockchain writing suggest that the system offers guarantees at most like \smartchain weak persistence.

Tendermint is another notorious permissioned platform that implements a variant of the PBFT protocol \cite{Buchman2016}, making its design more similar to \smartchain than Fabric.
However, Tendermint has two distinguished features: it uses a gossip protocol to propagate transactions among nodes, and it adopts a leader rotation mechanism similar to Spinning~\cite{Ver09}.
In terms of persistence, Tendermint writes the block before and after operation execution, making it less efficient than \smartchain (as evidenced by our experimental results), without further coordination between the replicas.
The consequence is that the system supports only weak persistence for its blockchain.

\paragraph{Consortium reconfiguration}
Some works have also tackled the challenges of supporting group reconfiguration in SMR~\cite{Lorch2006,Rodrigues2012,Ongaro2014,Pass2017,abraham2017}.
ComChain~\cite{Vizier19}, Hybrid Consensus~\cite{Pass2017}, and Solida~\cite{abraham2017} are the ones that most resemble our solution since they support fully autonomous reconfiguration.
Similarly to our approach, ComChain allows reconfigurations to be defined by application-specific criteria but does not deal with forks.
Hybrid Consensus determines the committee members using Bitcoin's (PoW-based) protocol while using a traditional consensus protocol among current committee members to order  transactions.
On the other hand, our solution is entirely derived from a classic BFT state machine protocol.
Moreover, Solida is designed to operate in the synchronous system model and uses a variant of the PBFT protocol adapted to such model.
Our solution is still able to operate in an eventually synchronous model, like most SMR protocols in the literature.

Fabric and Tendermint also support consortium reconfigurations.
Fabric only allows reconfiguration with the help of a trusted network administrator~\cite{FabricReconfig}. Tendermint, in principle, supports decentralized reconfigurations if the application defines how this should be done~\cite{TendermintReconfig}. 
However, none of these systems deal with the potential forks that might arise with multiple reconfigurations.


\section{Conclusions}
\label{sec:conclusions}

This paper discussed some misalignments between the state machine replication approach and the permissioned blockchain requirements and proposed several techniques to address them.
The identified issues concern the low performance of blockchain applications, the lack of strong blockchain persistence guarantees, and the possibility of forks due to consortium reconfigurations.
We propose a set of consensus-agnostic techniques materialized in a blockchain layer that can be integrated into SMR frameworks to mitigate these issues.
To validate our approach, we implemented these techniques on \smartchain, a proof-of-concept permissioned blockchain on top of \smart.
Experimental results show that \smartchain improves the performance of a simple digital coin application by $8\times$ when compared with running it on top of \smart, and by $8\times$ and $33\times$ when compared with Tendermint and Hyperledger~Fabric, respectively.


\subsubsection*{Acknowledgements} 
We thank Michael Davidson, Vincent Gramoli, Dragos-Adrian Seredinschi, the anonymous reviewers, and our shepherd, Heming Cui, for their comments that significantly improved this paper.
This work was supported by FCT through projects IRCoC (PTDC/EEI-SCR/6970/2014), ThreatAdapt (FCT-FNR/0002/2018), and the LASIGE Research Unit (UIDB/00408/2020 and UIDP/00408/2020), by FAPDF/Brazil through Edital 05/2018, and by the Swiss National Science Foundation (project number 175717).



\bibliographystyle{IEEEtran}
\bibliography{main}

\end{document}